\documentclass[twoside,12pt]{amsart}
\newtheorem{thm}{Theorem}[section]
\newtheorem{cor}[thm]{Corollary}
\newtheorem{lem}[thm]{Lemma}
\newtheorem{prop}[thm]{Proposition}

\makeatletter
\def\serieslogo@{}
\makeatother
\makeatletter
\def\@setcopyright{}
\makeatother

\theoremstyle{definition}

\theoremstyle{remark}
\newtheorem{rem}{Remark}[section]

\begin{document}

\title{An interesting class of \\
partial differential equations}

\author{Wen-An Yong}

\address{Zhou Pei-Yuan Center for Appl. Math.\\
Tsinghua University\\
Beijing 100084, China}


\email{wayong@tsinghua.edu.cn}



\begin{abstract}
This paper presents an observation that under reasonable 
conditions, many partial differential equations from 
mathematical physics possess three structural properties. 
One of them can be understand as a variant of the celebrated 
Onsager reciprocal relation in Modern Thermodynamics. It 
displays a direct relation of irreversible processes to the 
entropy change. We show that the properties imply various 
entropy dissipation conditions for hyperbolic relaxation 
problems. As an application of the observation, we propose 
an approximation method to solve relaxation problems. 
Moreover, the observation is interpreted physically and 
verified with eight (sets of) systems from different fields. 
\end{abstract}

\maketitle
\markboth{W.-A. Yong}{An interesting class of partial differential equations}

\tableofcontents

\section{Introduction}  

The goal of this paper is to draw attention to a class of 
partial differential equations (PDEs) of the form 
\begin{equation}\label{11}
U_t + \sum_{j = 1}^d F_j(U)_{x_j} = Q(U)  .
\end{equation}
Here $U$ is the unknown $n$-vector-valued function of 
$(x, t) \equiv(x_1, x_2, \cdots, x_d, t)\in{\bf R}^d
\times[0, +\infty)$, taking values in an open subset $G$ of 
${\bf R}^n$ (called {\sl state space}); $Q(U)$ and 
$F_j(U) (j = 1, 2, \cdots, d)$ are given $n$-vector-valued 
smooth functions of $U\in G$; and the subscripts $t$ and $x_j$ 
refer to the partial derivatives with respect to $t$ and $x_j$, 
respectively. 
  
As fundamental PDEs and as intermediate models \cite{Ga, Le, MR} 
between the Boltzmann equation \cite{Ce} and hyperbolic conservation 
laws \cite{Da}, systems of first-order PDEs with source terms 
describe various irreversible processes of scalar type \cite{GM}. 
Important examples occur in chemically reactive flows \cite{GiM}, 
radiation hydrodynamics \cite{MM, Po}, invisicid gas dynamics with 
relaxation \cite{Ze}, nonlinear optics \cite{HH}, 
and so on. 

Since the last decade, PDEs of the form (\ref{11}) have attracted 
much attention. See \cite{Na, Y3, Tz2} and references cited therein. 
One of the main interests is to identify a set of structural 
properties (axioms) that are satisfied by most of important 
equations from applications and, meanwhile, provide a convenient 
framework for the development of mathematical theories. In this 
regard, two stability conditions and various entropy dissipation 
conditions have been proposed in \cite{Y1} and 
\cite{CLL, MR, Y4, KY, Tz2}, respectively. See also 
\cite{Y3, RS}. All those conditions are generalizations of 
the well-known subcharacteristic condition \cite{Li} for 
(\ref{11}) with $n=2$ and $d=1$. For (\ref{11}), such a condition 
is the same in spirit as the H-theorem for the Boltzmann equation 
\cite{Ce} and as the entropy condition for conservation 
laws \cite{Da}.  

In this paper, we present an observation that under reasonable 
assumptions, many equations of the form (\ref{11}) from 
mathematical physics fall within a class characterized with the 
following three properties. (I) Every system in the class admits a 
strictly convex entropy function \cite{Go, FL}, 
(II) the source term can be written as a product of a non-positive 
symmetric matrix and the corresponding entropy variable, and (III) 
the symmetric matrix has a constant null-space. 

The first property is the well-known entropy condition for 
conservation laws and corresponds to the classical principles of 
thermodynamics. Property (II) can be understand as a 
variant of the celebrated Onsager reciprocal relation
in Modern Thermodynamics \cite{GM, KP} and implies 
the second law of thermodynamics. It displays a direct relation 
of irreversible processes to the entropy change. Property (III) 
expresses the fact that physical laws of conservation hold true, 
no matter what state the underlying thermodynamical system is in 
(equilibrium, non-equilibrium, and so on). 

We will verify the three properties for eight (sets of) systems 
of the form (\ref{11}) arising in gas dynamics with damping or 
with relaxation, nonlinear optics, radiation hydrodynamics, 
chemical reactions, kinetic theories (both moment closure systems 
and discrete velocity models), and so on. Furthermore, we show 
that the properties ensure a uniquely defined Maxwellian and 
imply various entropy dissipation conditions in the literature 
for hyperbolic relaxation problems. Thus, all the general results 
in \cite{Y2, YZ, MZ, RS, Y4, KY, Tz2, DY} apply to the  
aforementioned fields. 

We notice that the examples in Sections 8 and 9 have a common 
kinetic origin. Moreover, we know from \cite{Gi} that the chemical 
systems in Section 7 have a similar origin. It would be interesting 
to include the radiative gas example of Section 6 in the same basket, 
with the idea that radiation is associated with particles (photons) 
collisions. In other words, we show that the kinetic theory 
yields the Onsager relation for gas mixtures. 

As an application of our observation, we propose an approximation 
method to solve the relaxation problems. The accuracy of the 
method is analysed for initial value problems with smooth initial 
data, by using the results in \cite{Y2}. In this analysis, 
an important ingredient is a continuation principle for 
hyperbolic singular limit problems (Lemma 9.1 in \cite{Y3} and the 
appendix in \cite{BY}). Further analysis and applications of the 
approximation method are desirable. 
 
Our discussions indicate that the above three properties have a 
solid basis, from both mathematical and physical points of view. 
Thus, it seems reasonable to take the properties as 
requirements in construction of new mathematical models for 
irreversible phenomena. 

The paper is organized as follows. In Section 2 we present the 
observation and discuss its mathematical consequences. Section 3 
is devoted to the approximation method. Some physical 
interpretations are given in Section 4. The rest of the paper deals 
with the eight (sets of) examples.  

\section{An Observation}
\setcounter{equation}{0}

The main contribution of this paper is the following observation 
that under reasonable assumptions, many partial differential equations (PDEs) of the form 
(\ref{11}) arising in mathematical physics admit the following 
structure: \\ 

(I). There is a strictly convex smooth function $\eta(U)$ such that 
$\eta_{UU}(U)F_{jU}(U)$ is symmetric for all $U\in G$ and all $j$. 

(II). There is a symmetric and non-negative definite matrix 
${\mathcal L}={\mathcal L}(U)$ such that
$$
Q(U) = -{\mathcal L}(U)\eta_U(U).
$$

(III). The null space of ${\mathcal L}(U)$ is independent of 
$U\in G$.\\ 

\noindent Throughout this paper, $\eta_U(U)$ should be understand 
as a column vector. We will use $*$ as a superscript to denote 
the transpose operator. 

Recall that (I) is the classical observation due to Godunov 
\cite{Go}, and Friedrichs and Lax \cite{FL} for conservation laws 
$$
U_t + \sum_{j = 1}^d F_j(U)_{x_j} = 0 .  
$$

In what follows, we point out several important conclusions of the 
above observation. First of all, we show 

\begin{prop}\label{p21}
The observation is invariant under linear transformations of the 
form  
$$
V=PU. 
$$
Here $P$ is a constant and invertible $n\times n$-matrix.
\end{prop}
 
\begin{proof}
Let $\eta=\eta(U)$ be the strictly convex function in the 
observation. We need to show that the observation holds true with 
${\tilde\eta}(V)=\eta(P^{-1}V)$ for 
\begin{align}\label{21}
V_t + \sum_j(PF_j(P^{-1}V))_{x_j}=PQ(P^{-1}V). 
\end{align}
To this end, we compute 
$$
\eta_U=P^*{\tilde\eta}_V\qquad\mbox{and}\qquad
\eta_{UU}=P^*{\tilde\eta}_{VV}P.
$$
The second equality indicates that ${\tilde\eta}_{VV}(V)$ is 
positive definite, since so is $\eta_{UU}(U)$. The latter is 
equivalent to the strict convexity of $\eta(U)$. Therefore, 
${\tilde\eta}(V)$ is strictly convex. Since
$$
{\tilde\eta}_{VV}(PF_j(P^{-1}V))_V 
= (P^{-*}\eta_{UU}P^{-1})(PF_{jU}P^{-1})
=P^{-*}\eta_{UU}F_{jU}P^{-1}
$$
is symmetric, (I) follows. Moreover, since 
$$
PQ(P^{-1}V) = - P{\mathcal L}\eta_U = -P{\mathcal L}P^*{\tilde\eta}_{V} 
$$
and $P$ is independent of $U$, (II) and (III) follow.
This completes the proof. 
\end{proof} 

Let $r$ be such a constant that $(n-r)$ is the dimension of the 
null space in (III). In Proposition \ref{p21}, we take $P$ to be 
such a matrix that its first $(n-r)$ rows span the null space. 
Then the next proposition becomes obvious. 

\begin{prop}\label{p22}
Under the observation, there is a constant invertible matrix $P$ 
and a symmetric positive definite $r\times r$-matrix 
${\it l}(U)$ such that 
$$
P{\mathcal L}(U)P^* = \mbox{diag}(0_{(n-r)\times(n-r)}, {\it l}(U)) .
$$ 
Here and below we denote by $0_X$ the origin of ${\bf R}^X$ 
to avoid possible confusions. 
\end{prop}

With $P$ from Proposition \ref{p22}, the source term in (\ref{21}) 
obviously has the form
$$
PQ(P^{-1}V)=\begin{pmatrix}
0_{n-r}\\
q(V)
\end{pmatrix}
$$
with $q(V)\in{\bf R}^r$. Accordingly, we introduce the partition  
$$
V=\begin{pmatrix}
u\\
v
\end{pmatrix}
$$
and rewrite (\ref{21}) as 
\begin{equation}\label{22}
\begin{pmatrix}u\\[2mm] v\end{pmatrix}_t 
+ \sum_{j = 1}^d\begin{pmatrix}f_j(u, v)\\[2mm] 
g_j(u, v)\end{pmatrix}_{x_j} = 
\begin{pmatrix}0_{n-r}\\[2mm]q(u, v)\end{pmatrix}.    
\end{equation}
Notice that 
\begin{align}\label{23}
q(u, v) = -{\it l}(u, v){\tilde\eta}_v(u, v). 
\end{align}

Because $P$ is invertible and constant, (\ref{22}) with (\ref{23}) 
is equivalent to (\ref{11}). Thus, one may assume that (\ref{11}) 
is already in the form (\ref{22}) with (\ref{23}). In particular, 
the symmetric matrix ${\mathcal L}$ in (II) has the block-diagonal 
form in Proposition \ref{p22} and ${\it l}(U)$ therein is symmetric 
and positive definite.

With the equivalent form (\ref{22}), we turn to the following 
theorem, which is related to the so-called equilibrium manifold 
$$
{\mathcal E}:=\{U\in G: Q(U)=0\}.
$$    

\begin{thm}\label{p23}
Assume that (\ref{11}) possess the three observed 
properties and the state space $G$ is convex. Then for every 
$U\in G$, there is at most one point in ${\mathcal E}$, say $M=M(U)$, 
such that $U-M(U)$ is in the orthogonal complement of the null space. 
Moreover, there are two positive functions $c(U)<C(U)$, defined in 
the domain of $M=M(U)$, such that 
\begin{align}\label{24}
c(U)|U - M(U)|\leq|Q(U)|\leq C(U)|U - M(U)|.
\end{align}
Here $|X|$ denotes the Euclid norm of matrix or vector $X$. 
\end{thm}

Note that $M(U)$ may be defined only in part of the state space. 
A prototype of $MU)$ is the Maxwellian in kinetic theory \cite{Ce}.  

\begin{proof}
Assume that for a certain $U\in G$, there are two points 
$M_1, M_2\in{\mathcal E}$ such that both $U-M_1$ and $U-M_2$ are 
in the orthogonal complement of the null space. Let $P$ be the 
invertible matrix in Proposition \ref{p22} and set 
$$
PU=(u, v), \qquad PM_1 = (u_1, v_1), \qquad PM_2=(u_2, v_2).
$$
Then we have $q(u_1, v_1)=0=q(u_2, v_2)$ and $u_1 =u=u_2$. 
Since $l(u, v)$ is positive definite, from (\ref{23}) it follows 
that $q(u, v)=0$ if and only if ${\tilde\eta}_v(u, v)=0$. Thus 
we have ${\tilde\eta}_v(u, v_1)=0={\tilde\eta}_v(u, v_2)$. This  
contradicts the strict convexity of ${\tilde\eta}$ and thereby 
proves the first part of the theorem. 

For the inequalities in (\ref{24}), we denote by $m(U)$ the 
last $r$ components of $M(U)$. Since ${\tilde\eta}={\tilde\eta}(U)$ 
is strictly convex, there is a symmetric and positive definite 
$r\times r$-matrix $a(U)$ such that 
$$
{\tilde\eta}_v(PU)= {\tilde\eta}_v(PU)-{\tilde\eta}_v(PM(U))
=a(U)(v-m(U)).
$$
Thus, from (\ref{23}) we have $q(PU) = -l(PU)a(U)(v-m(U))$. 
Now the inequalities in (\ref{24}) can be easily verified with 
\begin{align*}
c(U) =& \big(|P||P^{-1}||l^{-1}(PU)a^{-1}(U)|\big)^{-1}, \\
C(U) =& |P||P^{-1}||l(PU)a(U)|. 
\end{align*}
This completes the proof. 
\end{proof}

The following theorem provides connections of the observation to 
the existing entropy dissipation conditions in the literature for 
hyperbolic relaxation problems. 

\begin{thm}\label{p24}
Assume that (\ref{11}) possess the three 
observed properties. Then the following conclusions hold:

\begin{enumerate}
\item $\eta_U^*(U)Q(U)\leq - \lambda^{-1}(U)|Q(U)|^2$ \quad 
for all $U\in G$, where $\lambda(U)$ is the maximum eigenvaue 
of ${\mathcal L}(U)$. \\
 
\item $Q(U)=0$ if and only if $\eta_U^*(U)Q(U)=0$ if and only 
if $\eta_U(U)$ is in the null space.\\

\item $\eta_U^*(U_e)Q(U)=0$ \quad for any $U, U_e\in G$ with $U_e$ 
satisfying $Q(U_e)=0$. \\

\item For $U$ satisfying $Q(U)=0$, $Q_U(U)\eta^{-1}_{UU}(U)$ is 
symmetric and non-positive definite, and its null space 
coincides with that of ${\mathcal L}(U)$.\\

\item For the equivalent version (\ref{22}) with (\ref{23}), 
$r\times r$-matrix $q_v(u, v)$ is invertible for $(u, v)$ satisfying 
$q(u, v)=0$. 

\end{enumerate}
\end{thm}

\begin{proof}
(1). Since ${\mathcal L}(U)$ is symmetric and non-negative definite, 
it is clear that 
${\mathcal L}^2(U)\leq \lambda(U){\mathcal L}(U)$. Therefore we have
\begin{align*}
\eta^*_U(U) Q(U) = -\eta^*_U(U){\mathcal L}(U)\eta_U(U)
\leq -\lambda^{-1}|{\mathcal L}(U)\eta_U(U)|^2=
- \lambda^{-1}(U)|Q(U)|^2.
\end{align*}
Note that the symmetry of ${\mathcal L}(U)$ is crucial to the 
inequality. 

(2). From (1) it follows immediately that $Q(U)=0$ if and only if 
$\eta^*_U(U)Q(U)=0$. Since 
$\eta^*_U(U)Q(U)= -\eta^*_U(U){\mathcal L}(U)\eta_U(U)$ and 
${\mathcal L}(U)$ is symmetric, $\eta^*_U(U)Q(U)=0$ is obviously 
equivalent to that $\eta_U(U)$ is in the null space of 
${\mathcal L}(U)$. 
 
(3). It follows from (II) that for $U_e$ satisfying $Q(U_e)=0$, 
$\eta_U(U_e)$ is in the null space of ${\mathcal L}(U_e)$. Thanks 
to (III), $\eta_U(U_e)$ is in the null space of ${\mathcal L}(U)$
for any $U\in G$. Thus, we have 
\begin{align}\label{25}
{\mathcal L}(U)\eta_U(U_e)=0 
\end{align}
and thereby $\eta^*_U(U_e)Q(U)=0$ for any $U, U_e\in G$ with $U_e$ 
satisfying $Q(U_e)=0$. 

(4). From (II) and (\ref{25}) it follows that     
\begin{align}\label{26}
Q(U) = -{\mathcal L}(U)(\eta_U(U)- \eta_U(U_e)) . 
\end{align}
With this relation, it is clear that 
$$
Q_U(U_e)= - {\mathcal L}(U_e)\eta_{UU}(U_e). 
$$
Thus (4) becomes obvious.

(5). Recall the block-diagonal form of ${\mathcal L}(u, v)$ for 
(\ref{22}) with (\ref{23}). It follows from (2) that 
${\tilde\eta}_v(u, v)=0$ for $(u, v)$ satisfying $q(u, v)=0$. Thus, 
we compute from (\ref{23}) that at $(u, v)$ satisfying $q(u, v)=0$, 
$q_v(u, v)=-l(u, v){\tilde\eta}_{vv}(u, v)$. Note that  
${\tilde\eta}_{vv}(u, v)$ is positive definite, since 
${\tilde\eta}(u, v)$ is strictly convex. This, together with the 
positive definiteness of $l(u, v)$, implies the invertibility 
of $q_v(u, v)$. Hence the proof is complete.  
\end{proof}
 
By Theorem \ref{p24}, if a system of PDEs possesses the 
three observed properties, then it satisfies all the entropy 
dissipation conditions in \cite{CLL, MR, Y4, KY, Tz2}. In fact, 
all the existing conditions consist of (I) and some additional 
requirements. They are $\eta^*_U(U)Q(U)\leq 0$, which is implied 
by (1) of Theorem \ref{p24}, and (2) of Theorem \ref{p24} in 
\cite{CLL}. In \cite{MR}, there is only one additional requirement 
which is $\eta^*_U(U)Q(U)\leq 0$ --- the second law of 
thermodynamics. In \cite{Y4}, the additional requirements are (1) 
and (5) of Theorem \ref{p24}, while they are (2) and (4) in 
\cite{KY}. The entropy dissipation condition in \cite{Tz2} is that 
in \cite{Y4} with $|Q(U)|$ in (1) replaced by $|U-M(U)|$. For this 
see the inequalities in (\ref{24}). Moreover, it was shown in 
\cite{Y4} that the entropy dissipation condition therein implies 
the stability conditions in \cite{Y1, Y2}. Thus, all the general 
results in \cite{DY, KY, MZ, RS, Tz2, Y2, Y4, YZ} are valid for 
PDEs of the form (\ref{11}) which possess the three observed properties. 

\section{An Approximation Method}
\setcounter{equation}{0}

Consider (\ref{11}) with a small parameter 
$\epsilon>0$: 
\begin{align}\label{91}
U_t + \sum_jF_j(U)_{x_j}=Q(U)/\epsilon. 
\end{align} 
This is the so-called relaxation problem. Assume the three observed 
properties in the previous section hold for this relaxation system 
(\ref{91}). Then the scaled system can be rewritten as 
\begin{align*}
U_t + \sum_jF_j(U)_{x_j}=-{\mathcal L}(U)\eta_U(U)/\epsilon. 
\end{align*} 
As an application of our observation, we show in this section 
that the above relaxation system can be approximated with the 
following simplified system
\begin{equation}\label{92}
U_t + \sum_jF_j(U)_{x_j}=-{\mathcal L}(U_*)\eta_U(U)/\epsilon 
\end{equation} 
as $\epsilon$ tends to zero. Here $U_*\in G$ is arbitrarily fixed.  

To this end, we use the equivalent form (\ref{22}):
\begin{equation}\label{93}
\begin{pmatrix}u\\[2mm] v\end{pmatrix}_t 
+ \sum_{j = 1}^d\begin{pmatrix}f_j(u, v)\\[2mm] 
g_j(u, v)\end{pmatrix}_{x_j} = \cfrac{1}{\epsilon}
\begin{pmatrix}0_{n-r}\\[2mm] q(u, v)\end{pmatrix}.    
\end{equation}
Here 
$$
q(u, v)=-{\it l}(u, v){\tilde\eta}_v(u, v)
$$ 
as in (\ref{23}). It is not difficult to see that as $\epsilon$ 
goes to zero, the formal limit of solutions to (\ref{93}) solves 
the following so-called equilibrium system 
\begin{align*}
u_t + \sum_jf_j(u, v)_{x_j}=0, \qquad q(u, v) =0. 
\end{align*}
This system consists of differential and algebraic equations. 

Since $q(u, v)=-{\it l}(u, v){\tilde\eta}_v(u, v)$ and 
${\it l}(u, v)$ is positive definite, the equilibrium system 
is equivalent to 
\begin{align*}
u_t + \sum_jf_j(u, v)_{x_j}=0, \qquad {\tilde\eta}_v(u, v) =0. 
\end{align*}
This system is independent of ${\it l}(u, v)$. By Theorem 
\ref{p23}, the algebraic equations define $v$ as a unique 
function of $u$, say, $v=h(u)$. Here we assume that the 
domain of $h(u)$ is non-empty and open! Thus, the equilibrium 
system becomes 
\begin{align}\label{94}
u_t + \sum_jf_j(u, h(u))_{x_j}=0, \qquad v=h(u). 
\end{align}
It is remarkable that $h(u)$, and thereby the equilibrium system, 
is independent of ${\it l}(u, v)$!

As is pointed out in the previous section, relaxation system 
(\ref{91}) satisfies the stability conditions in \cite{Y1, Y2}, 
for it possesses the three observed properties. Thus, Theorems 6.1 
and 6.2 in \cite{Y2} apply here: For smooth initial data, there 
is a finite and $\epsilon$-independent time interval $[0, T]$ such 
that the initial value problem of (\ref{91}) has a unique smooth 
solution $U^\epsilon=U^\epsilon(x,t)$ defined for $t\in[0, T]$ and 
satisfying 
\begin{align}\label{95}
U^\epsilon = P^{-1}\begin{pmatrix}u\\[2mm] h(u)\end{pmatrix}
 + O(\epsilon) 
\end{align}
in a certain Sobolev space, as $\epsilon$ goes to zero. See 
\cite{Y2} for details. Here $u$ solves the corresponding 
initial value problem of the equilibrium system in (\ref{94}).  
In addition, we have assumed for simplicity that the initial 
data take values in equilibrium and thereby initial-layers do 
not appear. Recall from \cite{Y2} that the time interval 
$[0,T]$ is the life-span of the smooth solution $u$.  
  
Note that (\ref{94}) is also the equilibrium system for the 
corresponding equivalent version (\ref{22}) of the simplified 
system (\ref{92}). The latter obviously possesses the three 
observed properties. Thus, we see that with the same initial 
data, the simplified system has a unique smooth solution 
${\hat U}^\epsilon$ defined in the same time interval and 
having the same expansion
\begin{align}\label{96}
{\hat U}^\epsilon = P^{-1}\begin{pmatrix}u\\[2mm] h(u)\end{pmatrix}
 + O(\epsilon)  
\end{align}
as $\epsilon$ goes to zero. Here the key points are the same 
equilibrium system and the same time interval. The latter is 
attributed to a continuation principle for hyperbolic singular 
limit problems (Lemma 9.1 in \cite{Y3}, see also the appendix 
in \cite{BY}). Consequently, we see from (\ref{95}) and 
(\ref{96}) that
$$
U^\epsilon - {\hat U}^\epsilon = O(\epsilon),  
$$
in a certain Sobolev space, as $\epsilon$ goes to zero. 

In conclusion, we have shown that for small $\epsilon$, relaxation 
systems (\ref{91}) and (\ref{92}) are close to each other in a 
finite and $\epsilon$-independent time interval for initial value 
problems with smooth data. The above discussion suggests an 
approximation method to solve the original relaxation system 
(\ref{91}). Further analysis and applications of this approximation 
method are desirable. In particular, it would be interesting to 
study the closeness for specific systems in the regime of non-smooth 
solutions. 

\section{Physical Interpretations}
\setcounter{equation}{0}

In this section, we give some physical interpretations of the 
three observed properties in Section 2. Recall that 
for a thermodynamic system inside which $n$ irreversible 
processes occur, the infinitesimal entropy change $dS$ due to 
the processes can be expressed as a sum of two parts: 
$$
dS = d_eS + d_iS.
$$
Here $d_eS$ is the part supplied to the system by its 
surroundings, and $d_iS$ is that produced inside the system. 
It is well known (see, e.g., \cite{GM}) that $d_eS$ corresponds to 
the flux terms in (\ref{11}) and $d_iS$ to the source term. The 
second law of thermodynamics states that $d_iS$ is zero for 
reversible processes and positive for irreversible ones.

Based on our observation, (\ref{11}) can be rewritten as 
\begin{equation}\label{31}
U_t + \sum_{j = 1}^d F_j(U)_{x_j} = -{\mathcal L}(U)\eta_U(U).    
\end{equation}
This form relates irreversible processes 
directly to the entropy change $\eta_U$. 

Recall that the physical entropy $S$ is equal to $-\eta$ and its 
existence is guaranteed by the classical principles of 
thermodynamics \cite{GM, KP}. This explains why the classical 
observation (I) has a solid basis in thermodynamics. The Gibbs 
relation on the total differential of $\eta$ (or $S$), in this 
general level, reads as 
\begin{align}\label{32}
d\eta = \eta_U(U)\cdot dU, 
\end{align}
where the dot $``\cdot"$ between two vectors means the scalar 
product. The usual Gibbs relation 
$$
\theta dS=de + pd\big(\cfrac{1}{\rho}\big)+ \cdots 
$$
is a slight rearrangement of (\ref{32}). Here $\theta$ is the 
temperature, $e$ is the specific internal energy, $p$ is the 
pressure, $\rho$ is the density, and the dots come from other 
possible internal variables. 

Property (II) very much looks like the celebrated Onsager 
reciprocal relations in Non-equilibrium Thermodynamics 
\cite{GM, KP}, if one understands the source terms 
as irreversible fluxes and the entropy variables as thermodynamic 
forces or affinities. However, it is slightly different from the Onsager 
relation. Firstly, it seems new to choose the entropy variables, 
instead of their linear combinations, as thermodynamic forces. Secondly, 
unlike the Onsager relation, 
$$
Q(U) = -{\mathcal L}(U)\eta_U(U)
$$ 
is a nonlinear relation between $Q(U)$ and $\eta_U(U)$. In fact, 
the matrix ${\mathcal L}={\mathcal L}(U)$ depends on $U$. Because 
the entropy function is strictly convex, there is a one-to-one 
correspondence between $U$ and the entropy variable $\eta_U(U)$ 
(see \cite{FL} for a proof of this fact). Thus, ${\mathcal L}$ 
depends on the entropy variable, which plays the role of affinities 
here. 

By the way, it is well known (see, e.g., \cite{Pe}, page 125--126) 
that there are difficulties in choosing the thermodynamic forces 
and fluxes when applying the notion. Here we have proposed an 
unconventional but unambiguous choice of the couple. 

Furthermore, we recall (\ref{26}) and deduce that 
for any $U, U_e\in G$ with $U_e$ satisfying $Q(U_e)=0$, 
\begin{align*}
Q(U) = & -{\mathcal L}(U)(\eta_U(U) - \eta_U(U_e)) \\
= & - {\mathcal L}(U_e)(\eta_U(U) - \eta_U(U_e)) - 
({\mathcal L}(U) - {\mathcal L}(U_e))(\eta_U(U) - \eta_U(U_e)) \\
= & - {\mathcal L}(U_e)\eta_U(U) + O(|U-U_e|^2). 
\end{align*}
Neglecting the higher-order term, we obtain a linear relation 
$$
Q(U) = - {\mathcal L}(U_e)\eta_U(U)  
$$
between $Q(U)$ and $\eta_U(U)$. Because ${\mathcal L}(U_e)$ is 
symmetric, this is the Onsager reciprocal relation if one 
considers the source terms as irreversible fluxes and the 
entropy variables as affinities. 

As to Property (III), we recall the equivalent form (\ref{22}) of 
(\ref{11}). In (\ref{22}), the first $(n-r)$ equations 
represent $(n-r)$ conservation laws. Note that $r$ might not have 
been a constant without assuming (III). In other words, 
Property (III) expresses the fact that the physical laws of 
conservation hold true, no matter what state the underlying 
thermodynamical system is in (equilibrium, non-equilibrium, 
and so on). 

\section{Four Specific Examples}
\setcounter{equation}{0}

From this section on, we will verify the three observed 
properties in Section 2 for a number of systems of the form 
(\ref{11}) arising in applications. This section contains 
four comparatively simple examples. \\

{\bf Example 1}. Multi-dimensional Euler equations of gas 
dynamics with damping: 
\begin{align*}
\rho_t + \mbox{div} (\rho u) = & 0, \\
(\rho u)_t + \mbox{div}(\rho u\otimes u) + \nabla p(\rho) 
= & -\rho u. 
\end{align*}
As usual, $\rho=\rho(x, t)$ stands for the density and 
$u=u(x, t)$ is the velocity. This system is of the form 
(\ref{11}) with 
$U=\begin{pmatrix}\rho\\ \rho u\end{pmatrix}\in{\bf R}^{d+1}$. 

It is well known that function 
$$
\eta(U) = \cfrac{\rho |u|^2}{2} + \int^\rho\int^\tau 
\cfrac{p'(\sigma)}{\sigma}d\sigma d\tau. 
$$
is a strictly convex entropy for the above system in the 
classical sense (I). By computing $\eta_U(U)$, we see that 
$$
Q(U) = -\mbox{diag}(0, \rho I_d)\eta_U(U), 
$$
where $I_k$ is the unit matrix of order $k$. Thus, the 
properties (II) and (III) obviously hold with 
${\mathcal L}(U)= \mbox{diag}(0, \rho I_d)$ for $\rho>0$. \\

Next three examples all have the form (\ref{22}) with $r=1$. 
For such a system, if there is a function $\eta=\eta(u, v)$ 
satisfying 
Property (I), then the observation is obviously true with 
$$
{\mathcal L}(U)= -\cfrac{q(u, v)}{\eta_v(u, v)}
\mbox{diag}(0_{(n-r)\times(n-r)}, 1) 
$$
($0_X$ is the origin of ${\bf R}^X$), provided that 
\begin{align}\label{41}
\cfrac{q(u, v)}{\eta_v(u, v)}<0
\end{align}
for all $(u, v)$ under consideration. The inequality  
(\ref{41}) is a stability condition for the corresponding 
systems. \\ 

{\bf Example 2}. A 3-D quasilinear system for 
nonlinear optics: 
\begin{align*}
{\vec D}_t - \nabla\times{\vec B} & = 0, \\
{\vec B}_t + \nabla\times{\vec E} & = 0, \\
\chi_t & = |{\vec E}|^2 - \chi
\end{align*}
with ${\vec D}=(1 + \chi){\vec E}$. See \cite{HH} for an 
explanation of the equations above. The state space here is 
$G=\big\{({\vec D}, {\vec B}, \chi): 
{\vec D}\in{\bf R}^3, {\vec B}\in{\bf R}^3, \chi>0\big\}
\subset{\bf R}^7$. 

Set  
$$
U=\begin{pmatrix}{\vec D}\\
{\vec B}\\
\chi\end{pmatrix}.
$$
In \cite{HH}, Hanouzet and Huynh showed that function 
$$
\eta(U) \equiv (1+\chi)^{-1}|{\vec D}|^2 + |{\vec B}|^2 
+ \chi^2/2 
$$ 
is a strictly convex entropy in the classical sense (I) in order 
to study the corresponding 
relaxation limit of the above system. By computing $\eta_U(U)$, 
we see that 
$$
Q(U) = -\mbox{diag}(0_{6\times 6}, 1)\eta_U(U).
$$
Thus, the observation is true with 
${\mathcal L}(U)= \mbox{diag}(0_{6\times 6}, 1)$. \\

{\bf Example 3}. 1-D Euler equations of gas dynamics in 
vibrational non-equilibrium (in Lagrangian coordinates): 
\begin{align*}
\nu_t - u_x = & 0, \\
u_t + p_x = & 0, \\
(e + \cfrac{u^2}{2})_t + (pu)_x = & 0, \\
q_t = & \omega(\theta_1) - \omega(\theta_2).
\end{align*}
See \cite{Ze} for an explanation of the equations above.
 
For this system, we know from \cite{Ze} that there is a strictly 
convex function $\eta=\eta(\nu, u, e+\cfrac{u^2}{2}, q)$ such that 
Property (I) holds and 
$$
\eta_q(U) = \cfrac{1}{\theta_1} - \cfrac{1}{\theta_2}.
$$
Then we have 
$$
Q(U) = -\theta_1\theta_2
\cfrac{\omega(\theta_1) - \omega(\theta_2)}{\theta_1 - \theta_2}
\mbox{diag}(0_{3\times 3}, 1)\eta_U(U) .
$$ 
Thus, the observation is true with 
$$
{\mathcal L}(U)= \theta_1\theta_2
\cfrac{\omega(\theta_1) - \omega(\theta_2)}{\theta_1 - \theta_2}
\mbox{diag}(0_{3\times3}, 1),  
$$ 
for $\omega=\omega(\theta)$ is strictly increasing \cite{Ze}.\\

{\bf Example 4}. 1-D Euler equations for isothermal motions 
of a viscoelastic material (in Lagrangian coordinates): 
\begin{align*}
\nu_t - u_x = & 0, \\
u_t + p_x = & 0, \\
(p + E\nu)_t = & -p - g(\nu).
\end{align*}
See \cite{Tz1} for an explanation of the equations above.

For this system, we know from \cite{Tz1} that function 
$$
\eta(U) = u^2/2 -E\nu^2/2 - p\nu
-\int^{-p - E\nu}h^{-1}(\sigma)d\sigma,  
$$
is a strictly convex entropy in the classical sense (I). Here 
$h^{-1}$ is the inverse of $h(\nu) = g(\nu) - E\nu$, which 
exists under the so-called subcharacteristic condition 
\begin{align}\label{42}
0<g_{\nu}(\nu)< E.
\end{align}
Since 
$$
\eta_p(U) = h^{-1}(-p - E\nu) - \nu, 
$$
we have 
$$
Q(U) = -\cfrac{p  + g(\nu)}{h^{-1}(-p - E\nu) - \nu}
\mbox{diag}(0_{2\times 2}, 1)\eta_U(U) . 
$$ 
Thus, the observation is true with  
$$
{\mathcal L}(U)= \cfrac{p  + g(\nu)}
{h^{-1}(-p - E\nu) - h^{-1}(h(\nu))}
\mbox{diag}(0_{2\times 2}, 1),  
$$ 
for $h(\nu)=g(\nu) - E\nu$ is strictly decreasing \cite{Tz1} under 
the subcharacteristic condition (\ref{42}).

\section{Radiation Hydrodynamics}
\setcounter{equation}{0}

In this section, we consider discrete-ordinate models of the 
Euler equations for radiation hydrodynamics \cite{MM, Po}, 
which are of the form (\ref{11}) with 
\begin{align*}
U =\begin{pmatrix}\rho\\ 
\rho v_1\\ \rho v_2\\ \rho v_3\\ \rho E\\ I_1\\ \vdots\\ I_L
\end{pmatrix},\ \ \ 
F_j(U) =\begin{pmatrix}\rho v_j\\ 
\rho v_1v_j + \delta_{1j}p\\ \rho v_2v_j + \delta_{2j}p\\ 
\rho v_3v_j + \delta_{3j}p\\ 
\rho Ev_j + pv_j\\ \mu^1_jI_1\\ \vdots\\ 
\mu^L_jI_L\end{pmatrix}, \ \ \ 
Q(U) = \begin{pmatrix}0\\ 0\\ 0\\ 0\\
C\rho\sum_{l=1}^L(I_l - B(\theta))\\ - \rho (I_1 - B(\theta))
\\ \vdots \\ - \rho (I_L - B(\theta))
\end{pmatrix}.
\end{align*}
Here $\rho$ is the density, $v_j$ is the velocity in the $j^{th}$ 
direction, $E = e + |v|^2/2$ with $e$ the specific internal energy, 
$I_l$ is the radiation intensity in the direction 
$\mu^l=(\mu^l_1, \mu^l_2, \mu^l_3)$, $p=p(\rho, e)$ is the 
pressure, $\delta_{ij}$ is the standard Kronecker delta, $C$ is 
a positive constant, and $B=B(\theta)$ is the 
Planck function of temperature $\theta$. 

For this system, the state space is 
$(0, \infty)\times{\bf R}^3\times(0, \infty)^{L+1}$. 
Since the basic assumptions of radiation hydrodynamics are not 
valid at low temperatures, we restrict the temperature domain 
to $[\theta_0, \infty)$ with $\theta_0>0$ a constant. 

Recall that $B=B(\theta)>0$ is strictly increasing with 
respect to $\theta\geq\theta_0$. We denote by $b=b(y)$ the 
inverse function of $B(\theta)$, that is, 
\begin{align}\label{51}
\theta = b(B(\theta)), \ \  \ \ \ \ \ \forall \ \theta\geq\theta_0. 
\end{align}
Note that $b=b(y)$ is strictly increasing. Moreover, it is 
smooth if so is $B=B(\theta)$. 

Define
\begin{align}\label{52}
\eta(U) = -\rho s(\rho, e) - C\sum^L_{l=1}\int^{I_l}_{B(\theta_0)}\cfrac{dy}{b(y)} 
\end{align}
with $s=s(\rho, e)$ the specific entropy. It is 
straightforward to verify that this $\eta$ is strictly convex. 
Since the system is the classical Euler equations coupled 
weakly to $L$ linear transport equations, $\eta$ is obviously 
an entropy function for the system. Namely, Property (I) is 
verified. 

Note that 
$$
\eta_{\rho E}(U)=-\cfrac{1}{\theta}, \qquad \eta_{I_l}(U)
=-\cfrac{C}{b(I_l)}
$$
and set 
$$
\sigma_l:=\cfrac{I_l - B(\theta)}{\theta^{-1} - b^{-1}(I_l)} .  
$$
Then it is not difficult to see that 
$$
Q(U)= - {\mathcal L}(U)\eta_U(U)
$$
with
\begin{align}\label{53}
{\mathcal L}(U)=\rho\begin{pmatrix}
0_{4\times4}&0_{4\times 1}&0_{4\times 1}&0_{4\times 1}&
0_{4\times 1}&\cdots&0_{4\times 1}\\
0_{1\times 4}&C\sum_l\sigma_l&-\sigma_1&-\sigma_2&-\sigma_3&
\cdots&-\sigma_L\\
0_{1\times 4}&-\sigma_1&C^{-1}\sigma_1&0&0&\cdots&0\\
0_{1\times 4}&-\sigma_2&0&C^{-1}\sigma_2&0&\cdots&0\\
0_{1\times 4}&-\sigma_3&0&0&C^{-1}\sigma_3&\cdots&0\\
\vdots&\vdots&\vdots&\vdots&\vdots&\vdots&\vdots\\
\vdots&\vdots&\vdots&\vdots&\vdots&\vdots&\vdots\\
0_{1\times 4}&-\sigma_L&0&0&0&\cdots&C^{-1}\sigma_L
\end{pmatrix}.
\end{align}
Since $\sigma_l>0$ for all $l$, this ${\mathcal L}(U)$ is 
symmetric and non-negative. Moreover, its null space is 
$$
\mbox{span}\left\{e_1, e_2, e_3, e_4, e_5 
+ C\sum_{l\geq6}e_l\right\},
$$
which is independent of $U$. Here $e_k$ is the $k^{th}$ 
column of the unit matrix $I_{(L + 5)}$. Consequently, the 
properties (II) and (III) are also verified.  

\section{Chemically Reactive Flows}
\setcounter{equation}{0}
 
Most of this section is taken from \cite{GiM}, except the 
verification of the properties II and (III). For multi-component 
reactive flows, if we neglect external forces, diffusion of mass, 
heat conduction and viscosity, and but retain the chemical 
reactions, the flows are described with PDEs of form 
(\ref{11}), where 
\begin{align}\label{61}
U =\begin{pmatrix}\rho_1\\ \rho_2\\ \vdots\\ \rho_{n_s}\\ 
\rho v_1\\ \rho v_2\\ \rho v_3\\ \rho E\end{pmatrix},\ \ \ 
F_j(U) =\begin{pmatrix}\rho_1 v_j\\ \rho_2 v_j\\ \vdots\\ 
\rho_{n_s} v_j\\ \rho v_1v_j + \delta_{1j}p\\ 
\rho v_2v_j + \delta_{2j}p\\ \rho v_3v_j + \delta_{3j}p\\ 
\rho Ev_j + pv_j\end{pmatrix}, \ \ \ 
Q(U) = \begin{pmatrix}m_1\omega_1\\ m_2\omega_2\\ \vdots\\ 
m_{n_s}\omega_{n_s}\\ 0\\ 0\\ 0\\ 0\end{pmatrix}. 
\end{align}
Here $\rho_k$ is the density of the $k^{th}$ species, $n_s$ 
is the number of the species, $\rho=\sum_k\rho_k$ is the total 
density, $v_j$ is the mass averaged flow velocity in the 
$j^{th}$ direction, $E = e + |v|^2/2$ with $e$ the specific 
internal energy of 
the mixture, $p$ is the pressure, $\delta_{ij}$ is the 
standard Kronecker delta, $m_k$ is the molar mass of the 
$k^{th}$ species (known constants), and $\omega_k$ is the 
molar production rate of the $k^{th}$ species. The system of will be closed by specifying $e, p$ and $\omega_k$ 
as functions of the natural variable
\begin{align}\label{62}
Y=(\rho_1, \rho_2, \cdots, \rho_{n_s}, v_1, v_2, v_3, \theta)^*
\end{align}
with $\theta$ the absolute temperature. 

We will specify $\omega_k$ later. $p$ and $e$ are given as 
in \cite{GiM}. For $p$, we denote by $R_g$ the universal gas 
constant, write $r_k=R_g/m_k$ and then define 
\begin{align}\label{63}
p = \theta\sum_k r_k\rho_k.  
\end{align}
$e$ is taken as the weighted average of the specific internal 
energy $\epsilon_k$ of the $k^{th}$ species: 
\begin{align}\label{64}
\rho e = & \sum_k\rho_k\epsilon_k , 
\end{align} 
where 
\begin{align}\label{65}
\epsilon_k = & \epsilon_k^0 + \int_{\theta_0}^\theta c_{vk}(y)dy. 
\end{align} 
Here $\epsilon_k^0$ is the specific internal energy of the 
$k^{th}$ species at the reference temperature $\theta_0>0$, 
and $c_{vk}=c_{vk}(\theta)$ are given smooth functions of 
$\theta\in [\theta_0, \infty)$, denoting the specific heat at 
constant volume of the $k^{th}$ species and satisfying 
$\min_{k, \theta}\{c_{vk}(\theta)\}>0$. 


The state space for the natural variable $Y$ defined in (\ref{62}) 
is $(0, \infty)^{n_s}\times{\bf R}^3\times[\theta_0, \infty)$. 
For the conserved variable $U$ defined in (\ref{61}), it is 
\begin{align*}
G\equiv\left\{U\in{\bf R}^{n_s + 4}: U_k>0 \ \mbox{for}\ 
1\leq k\leq n_s \ \mbox{and} \ 
U_{n_s +4}>\phi(U_1, U_2, \cdots, U_{n_s+3})\right\} , 
\end{align*}
where 
$$
\phi(U_1, U_2, \cdots, U_{n_s+3})=
\cfrac{U^2_{n_s + 1} + U^2_{n_s + 2} + U^2_{n_s + 3}}{2\sum_{k\leq n_s}U_k} +
\sum_{k\leq n_s}U_k\epsilon_k^0. 
$$
Since $\phi$ is a convex function, the state space $G$ is convex. 

Introduce 
\begin{align}\label{66}
s_k(\rho_k, \theta)= s_k^0 + 
\int^\theta_{\theta_0}\cfrac{c_{vk}(y)}{y}dy 
- r_k\ln\left(\cfrac{\rho_k}{m_k}\right) , 
\end{align}
where $s^0_k$ is a constant, and define 
\begin{align}\label{67}
\eta(U) = - \sum_k\rho_k s_k(\rho_k, \theta). 
\end{align}

We show that this $\eta=\eta(U)$ is an entropy function in the 
classical sense (I). Since 
\begin{align*}
\rho E = \sum_k\rho_k \left(\epsilon_k^0 + 
\int_{\theta_0}^\theta c_{vk}(y)dy\right) + \cfrac{\rho|v|^2}{2} 
\end{align*}
due to (\ref{64}) and (\ref{65}), we compute to obtain  
\begin{align*}
\theta_U = \left(\sum_k\rho_kc_{vk}(\theta)\right)^{-1} 
\left(\cfrac{|v|^2}{2} - \epsilon_1, \cdots, 
\cfrac{|v|^2}{2} - \epsilon_{n_s}, -v_1, -v_2, -v_3, 1\right).
\end{align*}
Thus, it follows from (\ref{67}) and (\ref{66}) that  
\begin{equation}\label{68}
\eta_U = \cfrac{1}{\theta}\left(\mu_1 - \cfrac{|v|^2}{2}, 
\cdots, \mu_{n_s} - \cfrac{|v|^2}{2}, v_1, v_2, v_3, -1\right)^*, 
\end{equation}
where 
\begin{align}\label{69}
\mu_k = \epsilon_k + r_k\theta -s_k\theta 
\end{align}
denotes the chemical potential of the $k^{th}$ species. On 
the other hand, by the definitions of $U$ and $Y$ in 
(\ref{61}) and (\ref{62}), we compute 
\begin{align*}
\cfrac{\partial Y}{\partial U} 
=\begin{pmatrix}I_{n_s}& 0_{n_s\times 3}&0_{n_s\times 1}\\
-\rho^{-1}v(1, 1, \cdots, 1)&\rho^{-1}I_3&0_{3\times 1}\\
&\theta_U&\end{pmatrix} 
\end{align*}
with $v=(v_1, v_2, v_3)^*$, and thereby  
\begin{align*}
\eta_{UU}(U)=& \cfrac{\partial \eta_U}{\partial U} =
\cfrac{\partial \eta_U}{\partial Y}\cfrac{\partial Y}{\partial U}\\
= & \begin{pmatrix}
\left(\delta_{kl}r_l\rho^{-1}_k\right)_{n_s\times n_s}& 
-\theta^{-1}(1, 1,\cdots, 1)^*v^*& 
\theta^{-2}(\cfrac{|v|^2}{2} - \epsilon_k)\\
0_{3\times n_s}&\theta^{-1}I_3&-\theta^{-2}v\\
0_{1\times n_s}&0_{1\times 3}&\theta^{-2}
\end{pmatrix}\cfrac{\partial Y}{\partial U}\\
=&\left(\cfrac{\partial Y}{\partial U}\right)^*
\begin{pmatrix}\left(
\delta_{kl}r_l\rho^{-1}_k\right)_{n_s\times n_s}& 0_{n_s\times 3}
& 0_{n_s\times 1}\\
0_{3\times n_s}&\rho\theta^{-1} I_3&0_{3\times 1}\\
0_{1\times n_s}&0_{1\times 3}&\theta^{-2}\sum_k\rho_kc_{vk}(\theta)
\end{pmatrix}
\cfrac{\partial Y}{\partial U} >0.
\end{align*}
Therefore, $\eta(U)$ is strictly convex. Moreover, since 
$$
\sum_j\xi_jF_j(U) = (v\cdot\xi) U + \begin{pmatrix}0_{n_s\times1}\\ 
\xi\\
v\cdot\xi\end{pmatrix}p  
$$
due to (\ref{61}), we have 
\begin{align*}
\partial_U\left(\sum_j\xi_jF_j(U)\right) =
(v\cdot\xi) I_{n_s + 4} + 
\begin{pmatrix}\rho_1\\
\rho_2\\
\vdots\\
\rho_{n_s}\\
\rho v\\
\rho E + p
\end{pmatrix}
\partial_U(v\cdot\xi)
+ \begin{pmatrix}0_{n_s\times 1}\\
\xi\\
v\cdot\xi\end{pmatrix}p_U . 
\end{align*}
Therefore, it follows from (\ref{68}), (\ref{63}), (\ref{64}), 
(\ref{69}) and (\ref{67}) that  
\begin{align*}
\eta_U^*\partial_U\left(\sum_j\xi_jF_j(U)\right)=
(v\cdot\xi)\eta_U^* + \eta\partial_U(v\cdot\xi)=
\partial_U(v\cdot\xi\eta).
\end{align*}
Hence, $\eta(U)$ is a strictly convex entropy function 
for the system in the classical sense (I). 

Next we turn to specifying $\omega_k$ by following \cite{GiM} again. 
Let the system have $n_r$ reversible reactions for $n_s$ species:
$$
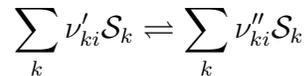
\sum_k\nu'_{ki}{\mathcal S}_k\rightleftharpoons
\sum_k\nu''_{ki}{\mathcal S}_k 
$$
for $i = 1, 2, \cdots, n_r$. Here ${\mathcal S}_k$ is the 
chemical symbol for the $k^{th}$ species, and $\nu'_{ki}$ 
and $\nu''_{ki}$ are the stoichiometric coefficients of the 
$k^{th}$ species in the $i^{th}$ reaction. The molar production 
rates $\omega_k$ are the Maxwellian production rates obtained 
in the kinetic framework of the ``slow reaction regime" or 
in the ``tempered reaction regime", when the chemical 
charactersitic times are larger than the mean free times 
of molecules:  
\begin{equation}\label{610}
\omega\equiv(\omega_1, \omega_2, \cdots, \omega_{n_s})^* = 
\sum_i\tau_i(\nu_{1i}, \nu_{2i}, \cdots, \nu_{n_si})^*
\equiv\sum_i\tau_i\nu_i . 
\end{equation}
Here $\nu_{ki} = \nu''_{ki} - \nu'_{ki}$ and $\tau_i$ is 
the rate of progress of the $i^{th}$ reaction: 
\begin{equation}\label{611}
\tau_i = K_{fi}(\theta)\prod_k
\left(\cfrac{\rho_k}{m_k}\right)^{\nu_{ki}'} 
- K_{ri}(\theta)\prod_k
\left(\cfrac{\rho_k}{m_k}\right)^{\nu''_{ki}} , 
\end{equation}
where $K_{fi}(\theta)$ and $K_{ri}(\theta)$ are the direct and 
reverse constants of the $i^{th}$ reaction, respectively; and 
\begin{equation}\label{612}
\cfrac{K_{fi}(\theta)}{K_{ri}(\theta)} = K_{ei}(\theta): = 
\exp\left(-\sum_k(r_k\theta)^{-1}\nu_{ki}\mu_k(m_k, \theta)\right)
\end{equation} 
with $\mu_k(m_k, \theta)$ the chemical potential (\ref{69}) 
at the unit concentration: $\rho_k/m_k =1$. 


It is well know that 
\begin{align*}
\sum_km_k\omega_k = 0 . 
\end{align*}
In fact, let $n_e$ be the number of elements involved in the 
system and denote by $e_{kl}$ the number of the $l^{th}$ element 
in the $k^{th}$ species. We have the {\sl element conservation} 
relations 
\begin{align}\label{614}
\sum_k\nu'_{ki}\epsilon_{kl} = \sum_k\nu''_{ki}\epsilon_{kl}
\end{align}
for $i = 1, 2, \cdots, n_r$ and $l= 1, 2, \cdots, n_e$. 
On the other hand, the species molar mass $m_k$ is related to 
the elemental masses $a_l$ by the relation
\begin{align*}
m_k = \sum_la_l\epsilon_{kl}. 
\end{align*}
Hence it follows from (\ref{610}) and (\ref{614}) that 
\begin{align*}
\sum_km_k\omega_k = \sum_k\sum_{i, l}a_l\epsilon_{kl}\tau_i\nu_{ki}
= \sum_{i, l}a_l\tau_i\sum_k\epsilon_{kl}(\nu''_{ki} - \nu'_{ki})=0. 
\end{align*}
Similarly, we have 

\begin{equation}\label{615}
\sum_km_k\nu_{ki} = 0 . 
\end{equation}

To see the properties (II) and (III), we set 
$$
M = \mbox{diag}(m_1, m_2, \cdots, m_{n_s}), \qquad 
{\mathcal Y} = (R_g\theta)^{-1}(\mu_1, \mu_2, \cdots, \mu_{n_s})^*. 
$$
and 
$$
\Delta_i =
K_{fi}(\theta)\prod_k\left(\cfrac{\rho_k}{m_k}\right)^{\nu'_{ki}}
\int_0^1\exp(\sigma<{\mathcal Y}, M\nu_i>)d\sigma >0. 
$$
It follows from (\ref{611}), (\ref{612}), (\ref{69}), 
(\ref{66}), (\ref{610}) and (\ref{615}) that 
\begin{align*}
\tau_i = & - \Delta_i\nu_i^*M{\mathcal Y} \\
= & - (R_g\theta)^{-1}\Delta_i\nu_i^*M
\left(\mu_1 - \cfrac{|v|^2}{2}, \mu_2 - \cfrac{|v|^2}{2}, 
\cdots, \mu_{n_s} - \cfrac{|v|^2}{2}\right)^*. 
\end{align*}
Note that the reaction rates $\tau_i$ depend exponentially on 
the chemical affinities $\nu_i^*M{\mathcal Y}$, due to 
$$
\int_0^1\exp(\sigma<{\mathcal Y}, M\nu_i>)d\sigma =
\frac{\exp(<{\mathcal Y}, M\nu_i>)-1}{\nu_i^*M{\mathcal Y}} . 
$$
Moreover, we set 
$$
V= (\nu_1, \nu_2, \cdots, \nu_{n_r}) \ \ \ \ \
\mbox{and} \ \ \ \ \ 
\Delta=\mbox{diag}(\Delta_1, \Delta_2, \cdots, \Delta_{n_r}). 
$$ 
Then we deduce from (\ref{61}), (\ref{610}) and (\ref{68}) that 
$$
Q(U) = -R_g^{-1}\mbox{diag}(MV\Delta V^*M, 0_{4\times 4})\eta_U(U).  
$$
Since $MV$ is a constant matrix and $\Delta$ is positive definite, 
the null space of $MV\Delta V^*M$ is independent of $U$. Hence 
the observation is verified with 
$$
{\mathcal L}(U) 
= R_g^{-1}\mbox{diag}(MV\Delta V^*M, 0_{4\times 4}). 
$$

Finally, let us mention that some statements of Proposition 2.4 
are also discussed in \cite{GiM}.

\section{Moment Closure Systems}
\setcounter{equation}{0}

Moment closure systems in kinetic theories are PDEs 
of the form (\ref{11}). In this section we show that our 
observation holds for the exponentially based closure systems 
in \cite{Le} corresponding to the Boltzmann equation
\begin{equation}\label{71}
f_t + \xi\cdot\nabla_x f = 
\int_{(\omega, \xi')\in S^{d-1}\times{\bf R}^d}
(f_\star f_\star' - ff')B(\omega, \xi, \xi')d\omega d\xi'.
\end{equation}
Here $f=f(x, t, \xi) \geq 0$ denotes the kinetic density of 
particles at the position-time-velocity point 
$(x, t, \xi)\in{\bf R}^d\times{\bf R}_+\times{\bf R}^d$, the 
dot $``\cdot"$ between two vectors means the scalar product, 
$f_\star = f(x, t, \xi_\star), f_\star' = f(x, t, \xi_\star')$ 
and $f' = f(x, t, \xi')$ with 
$$
\xi_\star = \xi - \omega\cdot(\xi - \xi')\omega 
\ \ \ \ \ \mbox{and} \ \ \ \ \ 
\xi_\star' = \xi' + \omega\cdot(\xi - \xi')\omega,   
$$
$B = B(\omega, \xi, \xi')$ is the collision kernel which is 
positive almost everywhere in its domain 
$S^{d - 1}\times{\bf R}^d\times{\bf R}^d$, and $d\omega$ is 
the normalized measure on the unit sphere $S^{d - 1}$. 

First of all, we recall the celebrated identity (see \cite{Ce})
\begin{equation}\label{72}
4\int\phi(\xi)(f_\star f'_\star - ff')B d\omega d\xi'd\xi = 
\int(\phi + \phi' - \phi_\star - \phi'_\star)
(f_\star f'_\star - ff')B d\omega d\xi'd\xi
\end{equation}
for any continuous function $\phi = \phi(\xi)$. Here and below, 
the integrals are taken over the whole domain and we write 
$\phi' = \phi(\xi'), \phi_\star = \phi(\xi_\star)$ and 
$\phi_\star' = \phi(\xi_\star')$. The identity is a direct 
result of the following symmetry properties of 
$B(\omega, \xi, \xi')$: 
\begin{equation}\label{73}
B(\omega, \xi, \xi') = B(\omega, \xi', \xi) = 
B(\omega, \xi_\star, \xi'_\star).
\end{equation}
Clearly, the integral in (\ref{72}) is zero (independent of $f$) 
if $\phi + \phi' = \phi_\star + \phi_\star'$. It is well-known 
\cite{Ce} that  
\begin{equation}\label{74}
\phi + \phi' = \phi_\star + \phi_\star'
\ \ \ \mbox{if \ and\ only \ if}\ \ \    
\phi(\xi)\in\mbox{span}
\{1, \xi_1, \xi_2, \cdots, \xi_d, |\xi|^2\}.
\end{equation} 

Moment closure systems considered here are derived from the 
Boltzmann equation as follows. Let $n$ be a positive integer 
and give $n$ linearly independent continuous functions 
$c_k = c_k(\xi)$ of $\xi\in{\bf R}^d$ ($k = 1, 2, \cdots, n$). 
Multiplying (\ref{71}) with $c_k(\xi)$ and integrating the 
resulting equations with respect to $\xi\in{\bf R}^d$ leads to 
$n$ equations 
\begin{equation}\label{75}
\partial_t\int c_k f d\xi + \nabla_x\cdot\int\xi c_k f d\xi 
= \int c_k(f_\star f_\star' - ff')Bd\omega d\xi'd\xi . 
\end{equation}
Let $\alpha_k=\alpha_k(x, t) (k = 1, 2, \cdots, n)$ be $n$ 
unknown scalar functions of $(x, t)$. Substituting 
\begin{equation}\label{76}
f = f(x, t, \xi) = \exp\Big(\sum_{k=1}^nc_k(\xi)\alpha_k(x, t)\Big) 
\end{equation}
into (\ref{75}), we get $n$ first-order PDEs for the $n$ 
unknown $\alpha_k$. 

\begin{rem}
Traditionally, each $c_k$ is a polynomial of $\xi$ and 
$\int c_k f d\xi$ is called a {\sl moment}. Here we do not 
require the $c_k$'s to be polynomials. 
\end{rem}

To make clear that the moment closure systems in (\ref{75}) 
with (\ref{76}) are of the form (\ref{11}), we write 
$c(\xi)\alpha = \sum_{k=1}^nc_k(\xi)\alpha_k$ and introduce the 
following functions of $\alpha\in{\bf R}^n$:
\begin{equation}\label{77}
\begin{split}
{\bar\eta}(\alpha) = & \int \exp\big(c(\xi)\alpha\big) d\xi , \\
q_j(\alpha) = & \int\xi_j \exp\big(c(\xi)\alpha\big) d\xi , \\ 
{\mathcal Q}(\alpha) = & \begin{pmatrix}{\mathcal Q}_1\\{\mathcal Q}_2\\
\vdots\\{\mathcal Q}_n\end{pmatrix} = 
\int\begin{pmatrix}c_1\\c_2\\\vdots\\c_n\end{pmatrix}
\big(\exp(c_\star\alpha + c_\star'\alpha) - \exp(c'\alpha +
c\alpha)\big)Bd\omega d\xi'd\xi. 
\end{split}
\end{equation}
Here we have considered 
\begin{equation}\label{78}
f_\star f_\star' = \exp(c_\star\alpha + c_\star'\alpha)
\ \ \ \ \  \mbox{and} \ \ \ \ \ 
ff' = \exp(c\alpha + c'\alpha)
\end{equation}
thanks to the Ansatz in (\ref{76}). With such an $f$, we formally 
have
$$
{\bar\eta}_{\alpha_k}\equiv
\cfrac{\partial{\bar\eta}}{\partial\alpha_k}
=\int c_k f d\xi\ \ \ \ \ \mbox{and} \ \ \ q_{j\alpha_k}
\equiv\cfrac{\partial q_j}{\partial\alpha_k}=\int\xi_j c_k fd\xi .
$$ 
Thus (\ref{75}) can be rewritten as 
\begin{equation}\label{79}
\cfrac{\partial{\bar\eta}_\alpha(\alpha)}{\partial t} + 
\sum_j\cfrac{\partial q_{j\alpha}(\alpha)}{\partial x_j} = 
{\mathcal Q}(\alpha). 
\end{equation}

Assume that {\sl there is a convex open set 
${\mathcal G}\subset{\bf R}^n$ such that the functions in 
(\ref{77}) are well-defined and smooth for 
$\alpha\in{\mathcal G}$}. The existence of such a ${\mathcal G}$ 
depends on the choice of the $c_k$'s and will not be addressed 
here. The interested reader is referred to \cite{Le}. 

Define 
$$
U = {\bar\eta}_\alpha(\alpha) \ \ \ \ \ \mbox{and} \ \ \ \ \  
G ={\bar\eta}_\alpha({\mathcal G}).
$$
We show that, for any $U\in G$, there is a unique 
$\alpha\in{\mathcal G}$ such that $U = {\bar\eta}_\alpha(\alpha)$. 
In fact, since the $c_k$'s are linearly independent, the Hessian 
matrix ${\bar\eta}_{\alpha\alpha}(\alpha)$ is symmetric 
positive definite and thereby ${\bar\eta}(\alpha)$ is strictly 
convex. Then the strictly convex function 
$({\bar\eta}(\alpha) - \alpha^*U)$ of $\alpha\in{\mathcal G}$ 
takes its local minimum at those $\alpha$ satisfying 
$U = {\bar\eta}_\alpha(\alpha)$. Since ${\mathcal G}$ is convex, 
there is at most one such minimum point. Consequently, $U={\bar\eta}_\alpha(\alpha)$ has a global inverse 
$\alpha=\alpha(U)$ for $U\in G$ and $G$ is diffeomorphic to 
the convex open set ${\mathcal G}$. 
 
For $U\in G$, set 
$$
\eta(U): = \alpha^*(U)U - {\bar\eta}(\alpha(U)). 
$$
We see that the inverse function $\alpha(U)$ is equal to 
$\eta_U(U)$. Thus, with 
\begin{equation}\label{710}
F_j(U):= q_{j\alpha}\big(\eta_U(U)\big) \ \ \ \ \mbox{and} 
\ \ \ \ Q(U):= {\mathcal Q}\big(\eta_U(U)\big), 
\end{equation}
we arrive at the following system of PDEs: 
\begin{equation}\label{711}
\cfrac{\partial U}{\partial t} + 
\sum_j \cfrac{\partial F_j(U)}{\partial x_j} = Q(U). 
\end{equation}
In \cite{Le}, Levermore showed that 
$\eta(U)$ defined above is a strictly convex entropy function 
for (\ref{711}) in the classical sense (I). 

To verify the observed properties (II) and (III), we use 
(\ref{72}) and rewrite ${\mathcal Q}(\alpha)$ defined in 
(\ref{77}) as 
\begin{align*}
{\mathcal Q}(\alpha) = \cfrac{-1}{4}\int
\begin{pmatrix}
c_{1\star} + c'_{1\star} -  c'_1 - c_1\\
c_{2\star} + c'_{2\star} -  c'_2 - c_2\\\vdots\\
c_{n\star} + c'_{n\star} -  c'_n - c_n
\end{pmatrix}
\big(\exp(c_\star\alpha + c'_\star\alpha)-
\exp(c'\alpha +c\alpha)\big)Bd\omega d\xi'd\xi.
\end{align*}
Notice that
\begin{align*}
& \exp(c_\star\alpha + c'_\star\alpha) - \exp(c'\alpha +c\alpha)\\
= & \int_0^1\exp[\sigma(c_\star + c'_\star - c' - c) \alpha 
+ (c' +c)\alpha]d\sigma(c_\star + c'_\star - c' - c)\alpha. 
\end{align*}
We set 
$$
b=b(\alpha, \xi, \xi', \xi'_\star, \xi_\star)
=\int_0^1\exp[\sigma(c_\star + c'_\star - c' - c) \alpha 
+ (c' +c)\alpha] d\sigma 
$$
and define 
\begin{align}\label{712}
a_{ij}(\alpha)= \cfrac{1}{4}
\int(c_{i\star} + c'_{i\star} -  c'_i - c_i)bB
(c_{j\star} + c'_{j\star} -  c'_j - c_j)d\omega d\xi'd\xi . 
\end{align}
Thus, ${\mathcal L}(U)=[a_{ij}(\eta_U(U))]_{n\times n}$ is a 
symmetric matrix and  
$$
Q(U)={\mathcal Q}(\eta_U(U))= -{\mathcal L}(U)\eta_U(U). 
$$
Since $b$ and $B$ are both positive, it follows from (\ref{712}) 
that ${\mathcal L}(U)$ is non-negative. Moreover, the null 
space of ${\mathcal L}(U)$ is 
$$
\Big\{\alpha\in{\bf R}^n :
\int\big|\big(c(\xi) + c(\xi') - c(\xi_\star) -  
c(\xi_\star')\big)\alpha|^2 d\omega d\xi d\xi' = 0\Big\} ,
$$
which is independent of $U$. Hence the observation holds for the 
moment closure systems in \cite{Le}. 

\section{Discrete Velocity Models}
\setcounter{equation}{0}

In this section, we consider discrete velocity models in 
kinetic theories \cite{Ga}:
\begin{equation}\label{81}
f_{kt} + a(k)\cdot\nabla_xf_k = Q_k(U)
\end{equation}
for $k=1, 2, \cdots, n$. Here $f_k=f_k(x, t)$ denotes the mass 
density of gas particles with the constant velocity 
$a(k)\in{\bf R}^d$ at time $t$ and position $x$, $a(k)\cdot\nabla_x=\sum_{j=1}^da_j(k)\partial_{x_j}$, 
$U=(f_1, f_2, \cdots, f_n)^*$, and $Q_k(U)$ is the collision 
term given by 
\begin{equation}\label{82}
Q_k(U) = \sum_{ijl}(A_{ij}^{kl}f_if_j - A_{kl}^{ij}f_kf_l),   
\end{equation}
where the summation is taken over all 
$i, j, l\in\{1, 2, \cdots, n\}$ and the coefficients 
$A_{ij}^{kl}$ are non-negative constants satisfying
\begin{equation}\label{83}
A_{ij}^{kl}=A_{kl}^{ij}= A_{lk}^{ij}.  
\end{equation}
It is not difficult to deduce from these symmetry properties that  
\begin{equation}\label{84}
\sum_{k=1}^n\phi_kQ_k(U) = \cfrac{1}{4}
\sum_{ijkl}A_{ij}^{kl}(\phi_k + 
\phi_l - \phi_i - \phi_j)(f_if_j - f_kf_l) 
\end{equation}
Remark that (\ref{83}) and (\ref{84}) are analogous to the 
fundamental properties in (\ref{73}) and (\ref{72}) of the 
Boltzmann equation. 

Our aim here is to show that, in the state space

$$
G:=\big\{f_k>0: \ \  k=1, 2, \cdots, n\big\}\ni U, 
$$
the discrete velocity model (\ref{81})-(\ref{83}) admits 
our observation with the strictly convex function 
\begin{align}\label{entropy8}
\eta(U)=\sum_{k=1}^nf_k(\log f_k -1). 
\end{align}
The strict convexity of $\eta(U)$ is obvious. Since it does 
not contain any cross-term, $\eta(U)$ is an entropy function 
for the diagonal and semilinear system (\ref{81}).

To see the properties (II) and (III), we set
$$
b_{ij}^{kl}=b_{ij}^{kl}(U)=\int^1_0\exp
[\sigma(\log f_i + \log f_j - \log f_k - \log f_l) + \log f_k + 
\log f_l]d\sigma > 0,
$$
which obviously has the symmetry properties (\ref{83}). Then 
the source terms can be rewritten as 
\begin{equation}\label{85}
\begin{split}
Q_k(U) & = \sum_{ijl}A_{ij}^{kl}b^{kl}_{ij}
(\log f_i + \log f_j - \log f_k -\log f_l)\\
& =-\sum_ma_{km}\log f_m, 
\end{split}
\end{equation}
where
$$
a_{km}(U)=-\sum_{jl}A_{mj}^{kl}b^{kl}_{mj} 
- \sum_{il}A_{im}^{kl}b^{kl}_{im} 
+ \sum_{ij}A_{ij}^{km}b^{km}_{ij} 
+ \delta_{km}\sum_{ijl}A_{ij}^{kl}b^{kl}_{ij} 
$$
with $\delta_{km}$ the Kronecker delta. Therefore, we have 
\begin{align}\label{86}
Q(U)= - [a_{km}]_{n\times n}\eta_U(U)
\equiv - {\mathcal L}(U)\eta_U(U), 
\end{align}
for $\eta_U(U) = (\log f_1, \log f_2, \cdots, \log f_n)^*$ due to 
(\ref{entropy8}). 

It remains to check the desired properties of ${\mathcal L}(U)$ 
defined in (\ref{86}). Thanks to the symmetry properties 
(\ref{83}) for both $b_{ij}^{kl}$ and $A_{ij}^{kl}$, it is not 
difficult to see that $a_{km} = a_{mk}$, that is, 
${\mathcal L}(U)$ is symmetric. Moreover, let 
$y=(y_1, y_2, \cdots, y_n)\in{\bf R}^n$. We refer to (\ref{84})
and (\ref{85}) to obtain 
$$
y{\mathcal L}(U)y^*= \cfrac{1}{4}
\sum_{ijkl}A_{ij}^{kl}b_{ij}^{kl}(y_i + y_j - y_k -
y_l)^2\geq 0. 
$$
Hence, ${\mathcal L}(U)$ is non-negative and its null space is 
$$
\big\{y\in{\bf R}^n: A_{ij}^{kl}(y_i + y_j - y_k -
y_l)=0 \quad \mbox{for \ all}\quad i, j, k, l\big\}.
$$
which is independent of $U$. Hence, our observation holds for 
the discrete velocity kinetic models constructed in \cite{Ga}. 

We conclude this paper by writing down the simplified 
system (\ref{92}) for the discrete velocity models. To do this, 
we compute from (\ref{entropy8}) that  
$\eta_U(U) = (\log f_1, \log f_2, \cdots, \log f_n)^*$. Then the 
corresponding simplified system reads as 
 
\begin{align*}
\begin{pmatrix}f_1\\[2mm]
f_2\\[2mm] \vdots\\[2mm]
f_n\end{pmatrix}_t + 
\sum_j\begin{pmatrix}a_j(1)&0&\cdots&0\\[2mm]
0&a_j(2)& \cdots&0\\[2mm]
\vdots&\vdots& \vdots&\vdots\\[2mm]
0&0&\cdots&a_j(n)\end{pmatrix}
\begin{pmatrix}f_1\\[2mm]
f_2\\[2mm] \vdots\\[2mm]
f_n\end{pmatrix}_{x_j}
= -{\mathcal L} \begin{pmatrix}\log f_1\\[2mm]
\log f_2\\[2mm] \vdots\\[2mm]
\log f_n\end{pmatrix}
\end{align*}
where ${\mathcal L}$ is a constant, symmetric and 
non-negative definite $n\times n$-matrix.

\end{document}